\documentclass[conference,a4paper]{IEEEtran}
\usepackage[pdftex]{graphicx}
\usepackage{amsmath,amssymb}
\usepackage{cuted}
\usepackage{cite}
\def\(({\left(}
\def\)){\right)}
\def\[[{\left[}
\def\]]{\right]}

\newcommand{\be}{\begin{equation}}
\newcommand{\ee}{\end{equation}}
\newcommand{\bea}{\begin{eqnarray}}
\newcommand{\eea}{\end{eqnarray}}

 \let\ee=\epsilon

\def\to{\rightarrow}

\newcommand{\beq}{\begin{equation}}
\newcommand{\eeq}{\end{equation}}

\begin{document}
\sloppy
%% Paper Title
%% You can use linebreaks \\ within to get better formatting as
%% desired. 
\title{Phase Transitions in Sparse PCA}
\author{
\IEEEauthorblockN{Thibault Lesieur} \IEEEauthorblockA{Institut de Physique Th\'eorique\\
    CEA Saclay and URA 2306, CNRS\\ 91191 Gif-sur-Yvette,
    France.}
\and \IEEEauthorblockN{Florent Krzakala} \IEEEauthorblockA{
Sorbonne Universit\'es, UPMC Univ. Paris 06\\
  LPS, CNRS UMR 8550 \\
  \'Ecole Normale Sup\'erieure, Rue Lhomond, Paris\\}
  \and \IEEEauthorblockN{Lenka
    Zdeborov\'a}  \IEEEauthorblockA{Institut de Physique Th\'eorique\\
    CEA Saclay and URA 2306, CNRS\\ 91191 Gif-sur-Yvette,
    France.}
}
%% To balance the two columns, you should reduce the text-height of
%% the last page using the following command:
%%%%%%%%%%%%%%%%%%%%%%%%%%%%%%%%%%%%%%%%%%%%%%%%%%%%%%%%%%%%%%%%%%%%%
%\addtolength{\textheight}{-9.35cm}
%%%%%%%%%%%%%%%%%%%%%%%%%%%%%%%%%%%%%%%%%%%%%%%%%%%%%%%%%%%%%%%%%%%%%
%% with an appropriate value. This command must be place on the second
%% last page, i.e., for a one-page abstract here, for a two-page
%% abstract right after the \maketitle command.
%% Create the title:
\maketitle
\begin{abstract}
    We study optimal estimation for sparse principal
    component analysis when the number of non-zero
    elements is small but on the same order as the dimension of the
    data. We employ approximate message passing (AMP) algorithm and its state
    evolution to analyze what is the information theoretically minimal
    mean-squared error and the one achieved by AMP in the limit of
    large sizes. For
    a special case of rank one and large enough density of non-zeros
    Deshpande and Montanari~\cite{deshpande2014information} proved
    that AMP is asymptotically optimal. We show that both for low density
    and for large rank the problem undergoes a series of phase transitions
    suggesting existence of a region of parameters where estimation is
    information theoretically possible, but AMP (and presumably every
    other polynomial algorithm) fails. The analysis of the large rank
    limit is particularly instructive. 
\end{abstract}

\section{Introduction}

Suppose we are given a data matrix $Y \in \mathbb{R}^{N\times N}$ that
was obtained from the following model 
\beq
      Y = \frac{1}{\sqrt{N}} X^T X + W \, ,\label{eq:model}
\eeq
where $X$ is a matrix in $\mathbb{R}^{r\times N}$. Each of the $N$ elements of $X$
is an independent random variable in $\mathbb{R}^{r}$ distributed according to
$P_0(x)$. Further $W \in \mathbb{R}^{N\times N}$ is a symmetric noise matrix
with elements $w_{\mu \nu}$ distributed as ${\cal N}(0,\Delta)$. 
The observation thus consists of a rank $r$ matrix corrupted by
a Gaussian noise. The main difficulty, but also interest, 
stems from the fact that we require $X$ to be sparse: only a fraction $\rho$
of elements of $X$ are non-zero and $P_0(x)$ is constrained 
accordingly. 

Let us denote by $X_0$ the true underlying signal matrix.  
We treat the problem of estimating $X$ from $Y$. We 
evaluate and analyze the estimator $\hat X$ that minimizes the mean
squared error 
\beq
    {\rm MSE} = \frac{1}{N}  \mathbb{E} \left( ||\hat X -X_0 ||^2_F \right)
    \, .
\eeq
If the distribution $P_0(x)$ is known then such an estimator is
given by the mean of the marginals of the posterior probability
distribution $P(X|Y)$. We analyze optimal estimation in the above
model in the limit where the system size $N$ is large, but rank is small $r\!=\!O(1)$, noise variance
$\Delta=O(1)$, and fraction of non-zero elements $\rho=O(1)$. 

We aim to answer the following two questions: (Q1) What is the
information theoretically minimal mean-squared error (MMSE) in this
limit? (Q2) In what range of parameters can the MMSE be achieved with
a computationally tractable algorithm? We answer these question by studying the approximate message passing (AMP) algorithm to estimate
the marginals of the posterior likelihood, and its asymptotic state
evolution (SE). Our results rely on a conjecture from statistical
physics, that the present problem belongs to a class of problems for
which the fixed points of the state evolution describe asymptotically
exactly both the optimal estimator and the performance of
AMP. Moreover current experience from other such problems suggests that when AMP
does not reach the MMSE then no other polynomial algorithm will.

\subsection{Motivation and background}

Principal component analysis (PCA) is a common dimensionality reduction
technique that aims at describing the data as linear combination of a
small number of principal components. In sparse PCA \cite{zou2006sparse} we search for
principal components with many zero elements to facilitate the
interpretation of the result. Only the non-zero elements then
correspond to features relevant for describing the variability in the
data. The sparsity constraint makes the problem
algorithmically challenging. Let us note that when talking about sparse PCA we have more often in
mind a model of a type $Y= U V^T + W$, rather than~(\ref{eq:model}). For simplicity of presentation, in this short
report, we restrict to the model (\ref{eq:model}), but applying
our method to the $UV^T$ setting is straightforward and leads to
comparable results.

Abundance of applications of sparse PCA
motivated both algorithmic development and theoretical studies of the
problem, see
e.g. \cite{zou2006sparse,moghaddam2005spectral,johnstone2009consistency,d2008optimal,krauthgamer2013semidefinite,deshpande2013sparse}. 
Many of these existing works are concerned with 
exact recovery of the support of non-zero elements. However, exact
recovery of the support is possible only when the number of non-zeros
is subextensive, i.e. the fraction of non-zeros $\rho = o(1)$ \cite{amini2009high}. In our setting, we assume
$\rho =O(1)$ which is reasonable in many applications. In this paper we model a typical case of sparse PCA by 
(\ref{eq:model}) and analyze this model in Bayesian probabilistic setting where we
assume the knowledge of the distribution $P_0(x)$. 

For sparse PCA of rank one $r=1$ the AMP algorithm and its state
evolution have been derived by Rangan and Fletcher
\cite{rangan2012iterative}. Deshpande and Montanari
\cite{deshpande2014information} were able to prove that for Bernoulli
distributed coordinates
(and always rank one) the state evolution equation indeed describes exactly the evolution of the
algorithm at large sizes when the density of non-zeros $\rho$ is large
enough. Remarkably, they also proved that the asymptotic MSE achieved by
AMP is in this case information theoretically optimal. Additionally,
in the regime where their proof is valid they did not
observe any phase transition in the MMSE. The corresponding AMP algorithm for generic rank was derived in
\cite{matsushita2013low}, however, without the state evolution. 

The result of Deshpande and Montanari about asymptotic optimality of
AMP is surprising for at least two reasons: First, for the question of
support recovery, there is a well
known large gap between what is information theoretically possible and
what is tractable with current algorithms \cite{amini2009high}. Moreover this gap was
linked to the problem of planted clique \cite{berthet2013computational}, where it is believed that no
polynomial algorithm will be able to achieve the information theoretic
performance. Second, in the regime where the rank is small but scales
linearly with $N$ an analogous gap between information theoretic and
tractable algorithmic performance was predicted to exist
\cite{kabashima2014phase}. This motivates us to revisit the analysis of the model
(\ref{eq:model}) in particular in the region of small density $\rho$
and for rank larger than one. Indeed, in both these cases we identify  phase transitions in
the MMSE, as well as regions where AMP is suboptimal. Technique-wise the
contribution of the present paper is a generalization of the
state evolution to arbitrary rank $r$, and analysis of the AMP-MSE and the MMSE for a
range of distributions $P_0(x)$. 

\section{From AMP to state evolution}

\subsection{AMP}

Here we remind the approximate message passing algorithm for general
rank $r$ \cite{matsushita2013low}, and sketch
the derivation of the corresponding state evolution. For rank one our
results reduce to the state evolution of
\cite{rangan2012iterative,deshpande2014information}. 

AMP is a large-$N$ simplification of the belief propagation equations (BP) \cite{YedidiaFreeman03} for a
graphical model corresponding to the posterior probability 
\beq
     P(X|Y) =   \frac{1}{Z(Y)} \prod_\mu  P(x_\mu)  
      \prod_{\mu \le \nu} \frac{e^{-\frac{(y_{\mu\nu} - x^T_\mu x_\nu/\sqrt{N})^2}{2\Delta}} }{\sqrt{2\pi \Delta}}
      \, . \label{eq:post}
\eeq
In the derivation we distinguish between $P(x)\neq P_0(x)$, but later we assume equality of the two.  
Variables in this graphical model are the $r$-component vectors
$x_\mu$ for $\mu = 1,\dots,N$. Belief propagation is written as an
iterative procedure on message that are probability distributions over
$x_\mu$. 

Remarkably the large-$N$ expansion of the corresponding BP
equations closes on messages $a^t_{\mu \to \nu} \in \mathbb{R}^{r}$ that are means of
the BP messages. 
%The resulting iterative equations read
%\begin{equation}
%a_{\mu \rightarrow \nu}^{t+1} = f(A^{t}_{\mu \rightarrow
%  \nu},B^{t}_{\mu \rightarrow \nu})\, ,
%\label{ComputeNewMean}
%\end{equation}%
%where the arguments are from $\mathbb{R}^{r\times r}$ and $\mathbb{R}^{r}$ respectively 
%\bea
%A_{\mu \rightarrow \nu}^t &= \frac{1}{N \Delta} \sum_{\gamma \neq
 % \nu, \gamma  \neq \mu} {a^t_{\gamma  \rightarrow \mu}} {(a^t_{\gamma
 %   \rightarrow \mu })^{\top}}\, ,
%\label{rAMP_A} \\
%B_{\mu \rightarrow \nu}^t &= \frac{1}{\Delta\sqrt{N}}
%\sum\limits_{\gamma  \neq \nu, \gamma  \neq \mu} y_{\mu \gamma } {a^t_{\gamma  \rightarrow \mu}}
%\label{rAMP_B}\, .
%\eea
The AMP algorithm is a further simplification of the corresponding equations
where the fact that messages $a^t_{\mu \to \nu} $ depend only
weakly on one of the index $\nu$ is exploited leading to a so called Onsager
correction term. This leads to the following AMP iterative algorithm
that is easily amenable to implementation
\bea
a_{\mu}^{t+1} = f(A^{t},B^{t}_{\mu})\, , 
~~~~
v_{\mu}^{t+1} = \frac{\partial f}{\partial B}(A^{t},B^{t}_{\mu })\, ,
\eea 
where $a_\mu \in \mathbb{R}^{r}$ $v_\mu \in \mathbb{R}^{r\times r}$,
the arguments are from $\mathbb{R}^{r\times r}$ and
$\mathbb{R}^{r\times r}$ respectively, and are given by
\bea
    &A^t = \frac{1}{N \Delta} \sum_{\mu} {a^t_{\mu}} {(a^t_{\mu})^{\top}}
\label{AMP_A}\, , \\
&B_{\mu}^t = \frac{1}{\Delta\sqrt{N}}
\sum\limits_{\nu} y_{\mu \nu } {a^t_{\nu}}  -
\frac{1}{\Delta N} \left(\sum_{\nu } v^{t-1}_\nu  \right)
{a^{t-1}_{\mu}}\, .
\label{AMP_B}
\eea
The function $f(A,B) \in \mathbb{R}^{r}$ is defined as the mean of the normalized
probability distribution
\begin{equation}
{\cal M}(x,A,B) = \frac{1}{{\cal N}(A,B)}  P(x)e^{  -\frac{1}{2}
 {x^{\top} A x} + B^{\top} x} \, .  \label{def:f}
\end{equation}
The AMP equations are usually initialized in such a way that $a^{t=0}$
is the mean of the prior distribution $P(x)$ and $v^{t=0}$ its
variance. 

Finally the AMP approach also provides an approximation for the
log-likelihood $\phi = \log Z(Y)$, where $Z(Y)$ is the normalization
of~(\ref{eq:post}). This is related to the Bethe free energy \cite{YedidiaFreeman03} simplified
along the very same lines as BP was simplified into AMP. 
Given a fixed point of the AMP algorithm we compute the Bethe log-likelihood
as
\begin{equation}
\phi = \frac{1}{N} \sum\limits_{\mu} \log{\cal N}(A,B_\mu) - \frac{1}{2N}\sum\limits_{\mu}
\log(\tilde Z_{\mu})\, ,
\label{BetheFreeEnergy}
\end{equation}
where ${\cal N}(A,B)$ is the normalization from (\ref{def:f}) and 
\beq
\tilde Z_{\mu} = \frac{1}{\Delta \sqrt{N}} {\rm Tr}(a_{\mu} \sum_\nu y_{\mu \nu}
a_{\nu}^{\top}) - \frac{1}{2\Delta N} {\rm Tr}(a_\mu a_\mu^{\top} \sum_\nu
a_\nu a_\nu^{\top}) \, .\label{SingleVariableFreeEnergy_2}  
\eeq

\subsection{State evolution}

State evolution describes the behavior of the AMP algorithm along
iterations via two order parameters from $ \mathbb{R}^{r\times r}$
\bea
Q^t &\equiv  \frac{1}{N} \sum\limits_{\mu} a^t_{\mu}
(a^t_\mu)^{\top}\, ,
\\
M^t &\equiv \frac{1}{N} \sum\limits_{\mu} a^t_\mu (x_0)_\mu^{\top} \, .
\eea
The mean squared error is related to these parameters as 
\bea
          {\rm MSE} = {\rm Tr}[ \mathbb{E}_{P_0}(x_0 x_0^{\top}) - 2 M +
          Q]  \label{MSE_SE} \, .
\eea 
With this definition we have from (\ref{AMP_A}) $A^t = Q^t/\Delta$, and
from (\ref{AMP_B}) by using (\ref{eq:model}) to express $y_{\mu \nu}$ and
neglecting sub-leading order terms, we
derive that $B_\mu^t$ is a
random Gaussian variable with mean $(x_0)_\mu M^t / \Delta$ and variance
$Q^t/\Delta$. The above order parameters hence follow the state
evolution equations
\bea
%Q^{t+1} = \iint {\rm d}{x_0} P_0(x_0){\rm d}{W}P_W(W)
%f f^{\top}(\frac{Q^t}{\Delta},\frac{M^t x_0}{\Delta} + W)
&Q^{t+1} = \mathbb{E}_{P_0(x_0),P_W(W)}
\left[ {f{(\frac{Q^t}{\Delta},\frac{M^t}{\Delta}x_0 + W)}
    f^{\top}{(.,.)}} \right],
\label{MSE_Q}
\\
%M^{t+1} =\iint {\rm d}{x_0} P_0(x_0){\rm d}{W}P_W(W)
%(\frac{Q^t}{\Delta},\frac{M^t x_0}{\Delta} + W) x_0^{\top}
&M^{t+1} =\mathbb{E}_{P_0(x_0),P_W(W)}  \left[
{f(\frac{Q^t}{\Delta},\frac{M^t}{\Delta}  x_0+ W)} x_0^{\top}\right],
\label{MSE_M}
\eea
where $W$ is a $r$-variate Gaussian random variable with zero mean
and of covariance $Q^t/\Delta$, the arguments of $f$ and
$f^{\top}$ in (\ref{MSE_Q}) are the same.
The Bethe log-likelihood can then be evaluated from the fixed point of
the state evolution as
\bea
     \phi &= \mathbb{E}_{P_0(x_0),P_W(W)}  \left[ \log {\cal
         N}(\frac{Q}{\Delta},\frac{M x_0}{\Delta}+W) \right]
\nonumber \\
&- \frac{1}{2 \Delta} {\rm Tr}(M M^{\top}) + \frac{1}{4 \Delta} {\rm
  Tr}(Q Q^{\top}) \, . \label{Bethe_SE}
\eea
We recall that $P(x)$ from (\ref{eq:post}) appears in
(\ref{MSE_Q}-\ref{MSE_M}) via the definition of the function $f(A,B)$
in (\ref{def:f}). 

In this paper we work in the
so-called Bayes-optimal setting where we
assume $P_0(x)=P(x)$, the state evolution then simplifies
because $M^t=Q^t$ for all $t$. This is called the Nishimori condition
in statistical physics \cite{kabashima2014phase} and was also derived in
\cite{deshpande2014information}. The intuition behind this condition
is that in Bayes-optimal inference the ground true signal $X_0$ behaves
in exactly the same way as a random sample from the posterior
distribution and hence $Q$ that describes the overlap between two
randomly chosen samples is the same as $M$ that describes the overlap
beween a randomly chosen sample and $X_0$. 

\subsection{Statistical physics conjecture}

The belief propagation equations from which we derived the AMP and the
state evolution assume that certain correlations between incoming
messages are weak enough. In statistical physics this assumption is
widely accepted to hold for inference in the Bayes optimal setting on models such as
(\ref{eq:model}), that correspond to a fully connected factor graph
with weak interactions on factor nodes. This has been used in many
works, see e.g. a more detailed discussion in \cite{kabashima2014phase}, and it has been proven in a subset of cases, see
notably the closely related \cite{deshpande2014information} or \cite{donoho2012information}. 

Under the above assumption and in the limit of large $N$,
the MMSE can be computed from a fixed point of the state evolution
equations (\ref{MSE_Q}-\ref{MSE_M}) that has the largest
log-likelihood~(\ref{Bethe_SE}). And the AMP-MSE can be computed from
a stable fixed point of the state evolution that is reached iteratively from
initialization $Q^{t=0}=\epsilon$ for a very small $\epsilon$.  

\section{Analysis of the state evolution}

\subsection{Phase of undetectability}

A first observation we make about the general state
evolution (\ref{MSE_Q}-\ref{MSE_M}) is that $M=Q=0$ is a fixed point
if and only if the prior distribution $P(x)$ in (\ref{eq:post}) has a
zero mean. This trivial
fixed point corresponds to as large MSE (\ref{MSE_SE}) as if all we knew about the
signal $X$ was the distribution $P_0(x)$. 
In case $Q=M=0$ is the fixed point with maximum log-likelihood 
then the matrix $Y$ did not contain any sign of 
the low rank perturbation, the information was completely lost in the
noise, and we denote the signal $X$ as undetectable. 

Whenever the mean of $P(x)$ is nonzero and $P_0(x)=P(x)$ then for
large but finite $\Delta$ the state evolution has a fixed point with
MSE smaller than $\mathbb{E}_{P_0}(x x^{\top})$. This means that for
distributions with non-zero mean the observed matrix $Y$ always
contains additional (to the prior) information about the signal, in
that case we say that the signal in detectable. In
this sense the sparse PCA problem is harder for distributions having
zero mean.   

We now study the linear stability of the fixed point $Q = 0$, $M
= 0$ in the case where both $P(x)$ and $P_0(x)$ have zero mean. We expand the state evolution equations around the trivial
fixed point up to the first order in $Q$ and $M$. Looking at the
Taylor expansion of
$f(Q/\Delta,M x_0 /\Delta + W)$, it is only the term $W \partial_B f(0,0) $ that will matter for eq. (\ref{MSE_Q}), and $M
x_0 \partial_B f(0,0)/ \Delta $ that will matter for
eq. (\ref{MSE_M}). Realizing that  from definition (\ref{def:f})
$\partial_B f(0,0)$ is the covariance of the distribution $P(x)$ we get
\begin{equation}
 Q^{t+1} = \frac{1}{\Delta}  \Sigma Q^t  \Sigma\, ,
~~~
M^{t+1} = \frac{1}{\Delta}  \Sigma M^t \Sigma_0\, ,
\end{equation}
where $\Sigma$ and $\Sigma_0$ are respectively the covariance matrices of $P(x)$ and $P_0(x)$.

When $\Sigma = \Sigma_0$ the above linearization will converge away
from the trivial fixed point for $\Delta<\Delta_u$, with
\begin{equation}
\Delta_u = \max\{\lambda^2 , \lambda \in {\rm Spectrum}(\Sigma)\}\, ,
\label{CritereStabilite}
\end{equation}
and the linearization is a contraction for $\Delta>
\Delta_u$. 
For distributions of $X$ of zero mean, there is hence a phase transition in
the behavior of AMP-MSE at $\Delta_u$. 
Translated into the behavior of the iterative AMP algorithm, when
$\Delta> \Delta_u$ AMP will converge to a trivial fixed point
$a_\mu=0$ for all~$\mu$, and  to a fixed point of smaller MSE for $\Delta<
\Delta_u$. It is quite remarkable to notice that this stability
criteria (\ref{CritereStabilite}) of the trivial fixed point is
universal, in the sense that it does not depend on the details of the distributions $P_0(x)$
and $P(x)$, it only requires their means to be zero and their covariances
to agree. 

The phase transition at $\Delta_u$ (and its universality) remarkably
reminds us of a detectability/undetectability spectral phase transition known for
the canonical PCA \cite{hoyle2004principal,baik2005phase}. The
difference is that in our setting there is no such phase transition when
$P(x)$ has a non-zero mean.

\subsection{The Gauss-Bernoulli case}

A particularly interesting and in our opinion representative example of distribution $P(x)$ that we will
(among others) study in this paper is the $r$-variate Gauss-Bernoulli
\begin{equation}
P(x) = P_0(x) = (1-\rho)\delta(x) + \frac{\rho}{\sqrt{2
    \pi}}\exp\left(\frac{-{x}^2}{2}\right) \, ,
\label{Sparse_Density}
\end{equation}
here $x \in \mathbb{R}^{r}$ and $\delta(x)$ is a $r$-dimensional Dirac delta function. 

Using the
criteria (\ref{CritereStabilite}) we conclude that for $\Delta > \rho^2$
AMP (randomly initialized) will not be able to detect that matrix $Y$
was a noisy low-rank matrix. On the other hand for $\Delta < \rho^2$ AMP will converge to
a fixed point giving informative MSE. 

For rotationally invariant distributions such as (\ref{Sparse_Density}),
i.e. $P(x)=P(R x)$ where $R$ any orthogonal matrix, we argue that
the rotational symmetry is also preserved in the state evolution. 
The covariance of rotationally invariant distributions must
be proportional to an identity $\mathbb{E}_{P}(x x^{\top}) = \sigma_0
\mathbb{I}$, with $\sigma_0 \in \mathbb{R}$. At the same time the order parameter $Q^t$ plays the role of an estimator
of this covariance and hence we assume that for all $t$ we have $Q^t =
q^t \mathbb{I}$, with $q^t \in \mathbb{R}$.  Note that for
rotationally invariant $P(x)$ the signal $X$ can be estimated only up
to a rotation~$R$.  In what follows we will always assume minima over
all possible $R$, which is equivalent to assuming $M^t = m^t
\mathbb{I}$.  The state evolution for (\ref{Sparse_Density}) and
generic rank $r$ can then be written explicitly as 
\bea
q^{t+1} \!\!\!\! &=&\!\!\!\! \frac{\rho  q^t}{\Delta + q^t}
J_r\left[\frac{q^t}{\Delta},\frac{q^t}{\Delta} +
  \frac{({q^{t}})^2}{\Delta^2} \right]\, ,
  \label{SE_q}
\\
J_r(a,\tau)\!\!\!\! &=&\!\!\!\! \int\limits_{0}^{+\infty}
{\rm d}{u} P_r(u)
\!
\left\{
1
 + 
\frac{\tau u^2  [1-\hat \rho(a,\tau u^2) ]}{r(1+a)}
\right\}\! \hat \rho(a,\tau u^2)\, ,
\nonumber
\eea
 where we used the Nishimori condition $q^t=m^t$, and 
\begin{equation}
P_r(u) = \frac{1}{ (2 \pi)^{\frac{r}{2}} }
\exp\left(-\frac{u^2}{2}\right) S_r u^{r-1}\, ,
\label{DensityHigh_r}
\end{equation}
$S_r$ being the surface of a unit sphere in $r$ dimensions.
And where $\hat \rho(a,b)$ is an estimator of the probability that a
given vector component is nonzero.
% Lenka: Is this a correct interpretation "probability that a
% given vector component is nonzero"?
\begin{equation}
\hat \rho(a,b^2)\ = \frac{\rho}{(1-\rho)\exp\left(-\frac{b^2}{2(1 +
      a)}\right)(1 + a)^{\frac{r}{2}} + \rho} \, .  \label{eq:rho}
\end{equation}
\subsection{The limit of large rank $r$}

In this section we show that for the Gauss-Bernoulli signal~(\ref{Sparse_Density}) when the rank $r$ is sufficiently large the state
evolution has a fixed point at $q_r = \rho-\Delta +o_r(1)$ whenever
$\Delta < \rho$, and this fixed point has always larger
log-likelihood than the trivial fixed point at $q=0$. Moreover, we
derive that the probability that a given component of the support is
correctly discovered goes to one exponentially fast in $r$ when $r$
grows. This means,
on the one hand,
that at large $r$ the MMSE of the sparse PCA behaves exactly as if the support of the non-zero elements was known,
i.e. ${\rm MSE}=\rho$ for $\Delta > \rho$ and ${\rm MSE}=\Delta$
for $\Delta < \rho$ as in \cite{baik2005phase}. AMP, on
the other hand, will
reach the MMSE only for $\Delta < \rho^2$ and will give ${\rm
  MSE}=\rho$ otherwise. There is hence a wide hard region in sparse
PCA between $\rho^2 < \Delta < \rho$ when $r\to \infty$. 

To prove the above large-$r$ results we analyze the function $\hat \rho(a, \tau u^2)$.
From the definition (\ref{eq:rho}) we can rewrite
\begin{equation}
\hat \rho(a,\tau u^2)\ = \frac{\rho}{(1-\rho)\exp\left(-\frac{\tau (u^2 - r)}{2(1 + a)} + \frac{r K(a,\tau)}{2}\right)+ \rho}
\label{Estimator_RhoLimit}
\end{equation}
with
\begin{equation}
K(a,\tau) = \frac{-\tau}{1 + a} + \log(1 + a)\, .
\end{equation}
From (\ref{DensityHigh_r}) we get that $u^2 - r$ has zero mean and
variance~$2r$.  The
exponential in (\ref{Estimator_RhoLimit}) is hence at large $r$
dominated by the term $K(a,\sqrt{\tau})$. 
%When $K(a,\sqrt{\tau}) > 0$ then $\hat \rho(a,\tau u^2) \rightarrow 0$
%and when $K(a,\tau) < 0$ then $\hat \rho(a,\tau u^2) \rightarrow 1$.
From concavity of $\log$ we see that for all $q \neq 0$ we have
$K\left(\frac{q}{\Delta},\frac{q}{\Delta} + \frac{q^2}{\Delta^2}
\right) < 0$. 
Therefore one can substitute
$\hat \rho(a, \tau u^2) = 1 + O(e^{-c r})$, where $c\in \mathbb{R}^+$, into the state evolution. Finally
using (\ref{SE_q}) we get for $r \rightarrow +\infty$ and $q \neq 0$
an iteration $q^{t+1} = \rho  q^t/(\Delta + q^t)$.
When $\Delta < \rho$ this equations has a stable fixed point  given by
$q = \rho - \Delta$. For $\Delta > \rho$ the only stable fixed point is $q=0$.

For the distribution (\ref{Sparse_Density}) the log-likelihood  (\ref{Bethe_SE})
becomes 
\begin{equation}
\phi = \mathbb{E}_{P_r(u)} \{ \rho
  \psi[\frac{q}{\Delta},(q+\Delta)\frac{q u^2}{\Delta^2} ]
  + (1-\rho) \psi(\frac{q}{\Delta},\frac{q u^2}{\Delta}) \} -
  \frac{r q^2}{4 \Delta}  \,,  \nonumber
\end{equation}
where $\psi(a,\tau u^2 ) = \log[{\cal N}(a \mathbb{I}, \sqrt{\tau} u
\vec{e})]$ with $\vec{e}$ being an arbitrary vector of unit norm. 	

For the fixed point $q = 0$ we get $\phi=0$.
To compare to the log-likelihood of the fixed point 
$q=\rho-\Delta$ we develop $\psi(a,\tau u^2 )$ for large $r$ and
assume $q\neq 0$.  
For $K(a,\tau) < 0$ we have $\psi(a,\tau u^2 ) \approx \log(\rho)- \frac{r}{2}K(a,\tau)$,
from which we obtain 
\begin{equation}
\phi(q) = -\frac{\rho r}{2}\left[
\log\left(1 + \frac{q}{\Delta}\right) - \frac{q}{\Delta} + \frac{q^2}{2 \rho\Delta}
\right]+ O_r(1)\,,
\label{Large_R_FreeEnergy}
\end{equation}
which evaluated at $q = \rho - \Delta$ is positive (and hence larger
that the value for the fixed point $q=0$) if and only if $\Delta <
\rho$. This tells us that at large enough $r$ and  $\Delta <
\rho$ the fixed point $q = \rho - \Delta$ corresponds to the MMSE.

\begin{figure}
\centering
\includegraphics[scale = 0.55]{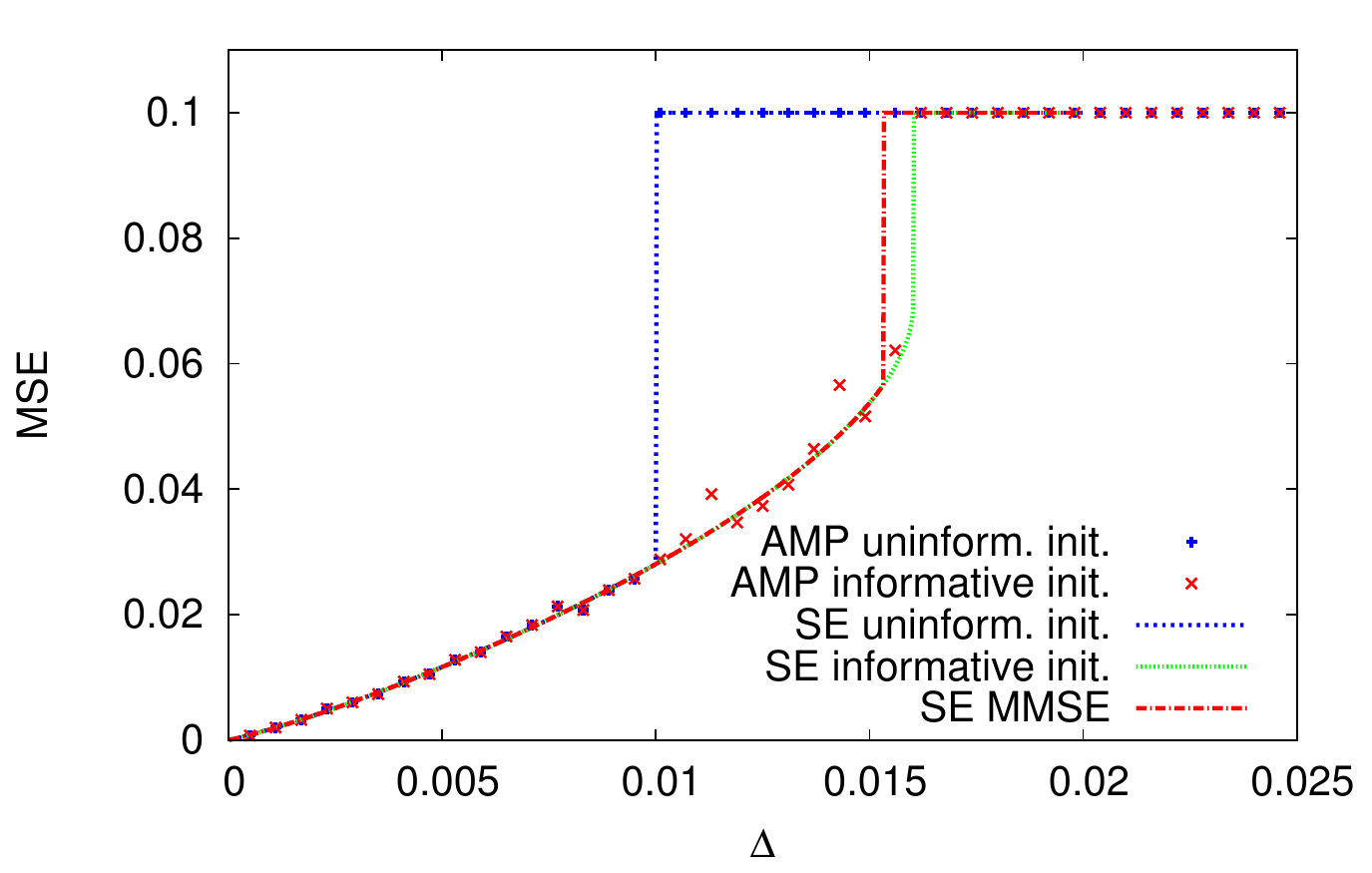}
\vspace{-4mm}
\caption{Example of the discontinuous phase transitions in the MSE for sparse PCA.
  The data are for the Gauss-Bernoulli distribution
  (\ref{Sparse_Density}), rank one and density $\rho=0.1$. The lines
  are results of the state evolution, the points of the AMP algorithm
  run on one random instance of the problem with $N=20000$. In blue
  (left most) is
  the MSE reached from an uninformative initialization of the
  SE/AMP. In green (right most) is the MSE reached from the informative
  initialization. In red (middle) is the MMSE. The discontinuities are at
$\Delta_{\rm AMP}=0.0100(1)$, $\Delta_c=0.0153(1)$, $\Delta_{2nd}=0.0161(1)$. 
%In black, corresponding to the right $y$-axes is the probability that a
%given non-zero element of the signal was not correctly identified.
}
\label{plot_MSE}
\vspace{-4mm}
\end{figure}

\subsection{Numerical Results}

In this section we give several examples of a numerical investigation
of the fixed points of the AMP algorithm and of the state evolution. 
In all these experiments we iterate the corresponding equations till
convergence and monitor the corresponding mean squared error and the
value of the log-likelihood. We initialize the iterations in two
different ways.
\begin{itemize}
   \item{Uninformative initialization: In state evolution this means
       $Q^{t=0}=\epsilon$, where $\epsilon$ is very small. In
     the AMP algorithm this corresponds to $a_\mu^{t=0}=\epsilon_\mu$
   and $v_\mu^{t=0}=\mathbb{E}_{P_0}(x_0 x_0^{\top})$.}
   \item{Informative initialization: In the state evolution this means
       $Q^{t=0} = \mathbb{E}_{P_0}(x_0 x_0^{\top})$, in AMP this means
     $a_\mu^{t=0}=(x_0)_\mu$ and $v_\mu^{t=0}=0$.}
\end{itemize}
In a region where these two initializations converge to a different
fixed point the MMSE is the one for which the
log-likelihood evaluated from (\ref{Bethe_SE}) is larger. In
Fig.~\ref{plot_MSE} we compare the MSE reached by the state evolution
and the AMP algorithm on an instance of size $N=20000$ and as expected
we see an excellent agreement.

\begin{figure}
\centering
\includegraphics[scale = 0.55]{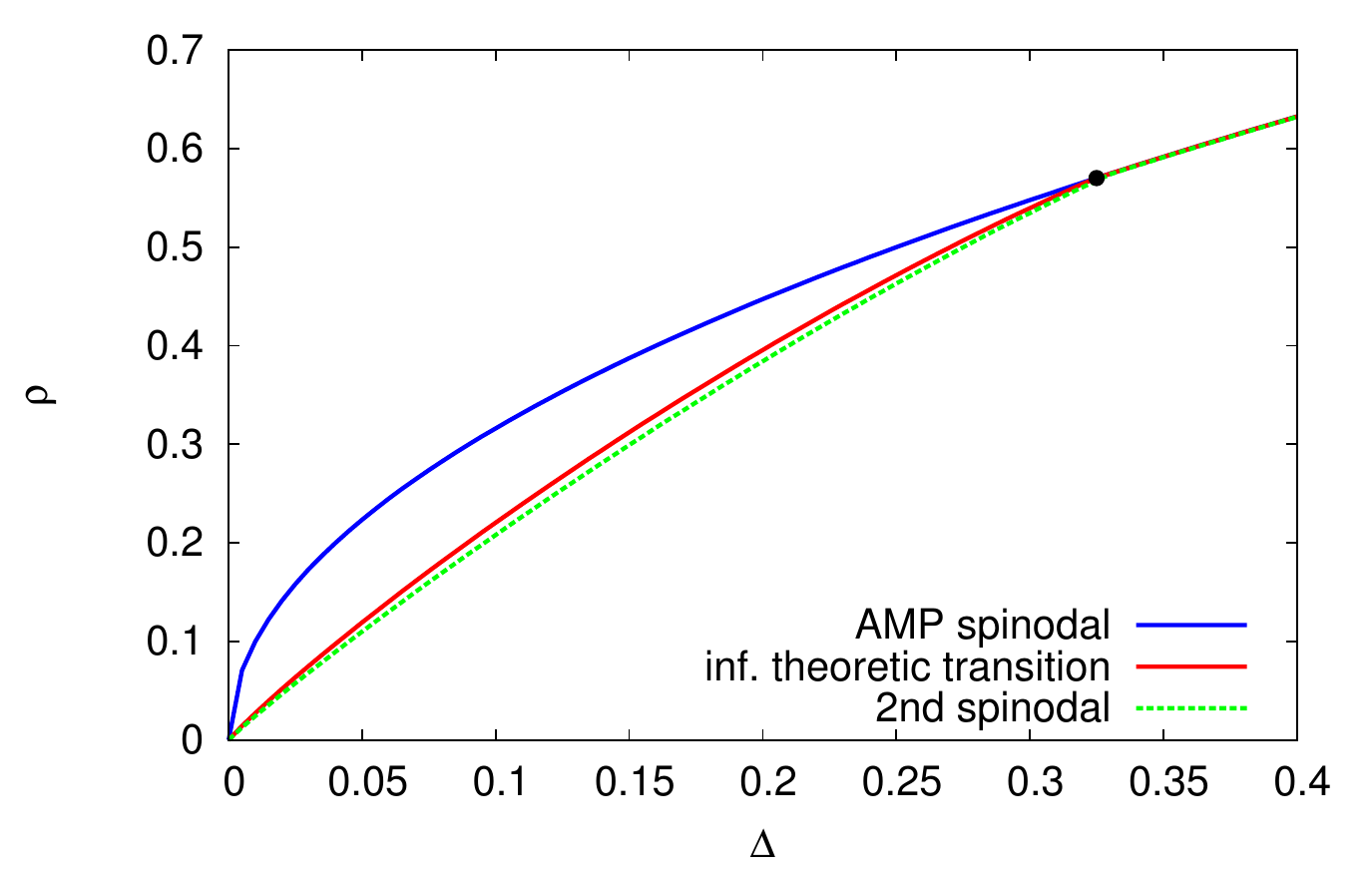}
\vspace{-4mm}
\caption{The phase diagram in the density $\rho$ versus noise $\Delta$
  plane for the Gauss-Bernoulli signal (\ref{Sparse_Density}), rank
  $r=50$. The curves from above are: In blue is the $\rho_{\rm AMP}$ above which
  AMP is asymptotically able to find the MMSE. In red is the phase
  transition in the MMSE $\rho_c$.
  %, for $\Delta<0.32$ this is a discontinuous
  %transition, for $\Delta>0.32$ a continuous one. 
  In green is the 2nd spinodal $\rho_{\rm 2nd}$ below which the
  informative fixed point does not exist anymore.
  For $\Delta>0.32(1)$ the transition is continuous and the three
  critical values are equal to $\rho_u=\sqrt{\Delta}$.}
\label{plot_large_r}
\vspace{-4mm}
\end{figure}

Figs. \ref{plot_MSE} and \ref{plot_large_r} are for the Gauss-Bernoulli signal
distribution for which the trivial fixed point exists and
is locally stable for $\Delta>\rho^2$. In this case of zero mean
$P_0(x)$ there is either a single second order (continuous) phase transition
at $\Delta_u=\rho^2$ or a (discontinuous) first order phase transition
in the MMSE at
$\Delta_c>\rho^2$ with its two spinodals, $\Delta_{\rm AMP}$
and $\Delta_{\rm 2nd}$. By the results of the previous section
in the limit of large rank we have $\Delta_c(r \to \infty)=\Delta_{\rm
  2nd}(r \to \infty)=\rho$.  Fig.~\ref{plot_large_r} depicts the result
for rank $r=50$. 

\begin{figure}
\centering
\includegraphics[scale = 0.55]{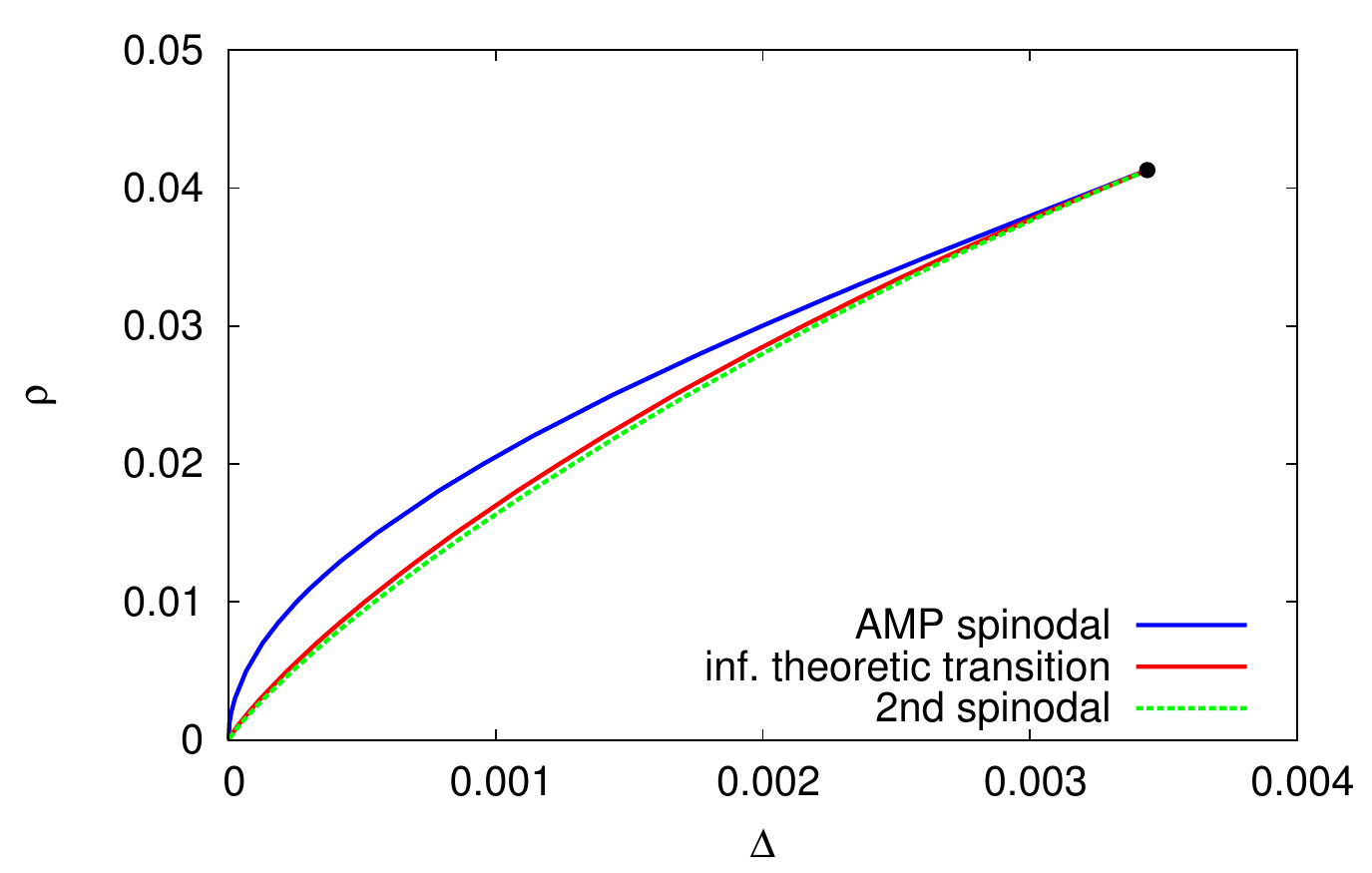}
\vspace{-4mm}
\caption{The phase diagram in the density $\rho$ versus noise $\Delta$
  plane for the rank-one spiked Wigner model investigated in
  \cite{deshpande2014information}, i.e. $P_0(x)=\rho \delta(x-1) + (1-\rho) \delta(x)$. This signal distribution does not
  have a zero mean and hence the trivial fixed point of the state
  evolution does not exist. This translates into the fact that for
  $\rho> 0.041(1)$ the AMP is information theoretically optimal and there
  is no phase transition. For low densities we, however, again observe
  the three phase transitions, from above: In
  blue is the AMP spinodal $\rho_{\rm AMP}$ above which
  AMP is asymptotically able to find the MMSE. In red is the
  discontinuous phase transition in the MMSE $\rho_c$. Therefore, AMP is suboptimal
  in the region between the blue and red curve, for
  $\rho_c<\rho<\rho_{\rm AMP}$. In green is the
  2nd spinodal $\rho_{\rm 2nd}$ below which the informative fixed point does not exist
  anymore.}
\label{plot_spike}
\vspace{-4mm}
\end{figure}

In Fig.~\ref{plot_spike} we depict the result for a case of
$P_0(x)$ with a non-zero mean. In that case either there is a unique
fixed point and no phase transition, or there are two fixed points,
both with MSE smaller than if the signal was chosen randomly only
according to the prior information. Remarkably, the first order phase
transition seems always to happen for small
densities. Fig.~\ref{plot_spike} depicts the case of spiked Wigner model
from \cite{deshpande2014information}, in a region of
densities for which the proof of \cite{deshpande2014information} did
not apply. Note that the AMP spinodal $\rho_{\rm AMP}$ and the curve of phase transition
in MMSE $\rho_c$ arrive with a different slope to $\rho\to 0$, this is
consistent with the algorithmic gap for support recovery for
sub-extensive support size \cite{amini2009high}.

\section{Conclusions}

In this paper we analyzed probabilistic estimation in sparse PCA
modeled by (\ref{eq:model}). We derived the state evolution of the
approximate message passing algorithm for general rank $r$. Relying on a
statistical physics conjecture about exactness of this state evolution
we analyze it's fixed points to compute the minimal mean squared error
and the error made by AMP in the large $N$ limit. Generalization of the proof technique  of
\cite{deshpande2014information} to small densities, ranks larger than
one and signal distributions having zero mean are an important topic
for future work. 

For signal distributions of zero mean we unveil an undetectability regime where no
algorithm can do better in estimation of the signal than random
guessing. We observe a first order
(discontinuous) phase transition in regions of small density $\rho$
or large rank $r$. Existence of such a first order phase transition is
related to the existence of a region of parameters where AMP is
asymptotically sub-optimal. 

In the large rank limit
the emerging picture is particularly simple (at least for signal
distributions of zero mean). The asymptotic MMSE is equal to the MMSE
of the problem with known support, and the probability of false
negatives in the support recovery is exponentially small when $r\to
\infty$. The MMSE is not reachable for AMP unless the
noise variance is smaller than the variance of the signal {\it squared}. 
We expect that this {\it hard} region will stay computationally hard
also for other polynomial algorithm. Proving such a result
about algorithmic barrier for a generic class of algorithms is a very interesting challenge.

\section*{Acknowledgment}
This work has been supported in part by the ERC under the European
Union’s 7th Framework Programme Grant Agreement 307087-SPARCS.

%% References:
%% We recommend the usage of BibTeX:
%%
\bibliographystyle{IEEEtran}
\bibliography{refs}

\begin{thebibliography}{10}
\providecommand{\url}[1]{#1}
\csname url@samestyle\endcsname
\providecommand{\newblock}{\relax}
\providecommand{\bibinfo}[2]{#2}
\providecommand{\BIBentrySTDinterwordspacing}{\spaceskip=0pt\relax}
\providecommand{\BIBentryALTinterwordstretchfactor}{4}
\providecommand{\BIBentryALTinterwordspacing}{\spaceskip=\fontdimen2\font plus
\BIBentryALTinterwordstretchfactor\fontdimen3\font minus
  \fontdimen4\font\relax}
\providecommand{\BIBforeignlanguage}[2]{{%
\expandafter\ifx\csname l@#1\endcsname\relax
\typeout{** WARNING: IEEEtran.bst: No hyphenation pattern has been}%
\typeout{** loaded for the language `#1'. Using the pattern for}%
\typeout{** the default language instead.}%
\else
\language=\csname l@#1\endcsname
\fi
#2}}
\providecommand{\BIBdecl}{\relax}
\BIBdecl

\bibitem{deshpande2014information}
Y.~Deshpande and A.~Montanari, ``Information-theoretically optimal sparse
  pca,'' in \emph{Information Theory (ISIT), 2014 IEEE International Symposium
  on}, 2014, pp. 2197--2201.

\bibitem{zou2006sparse}
H.~Zou, T.~Hastie, and R.~Tibshirani, ``Sparse principal component analysis,''
  \emph{Journal of computational and graphical statistics}, vol.~15, no.~2, pp.
  265--286, 2006.

\bibitem{moghaddam2005spectral}
B.~Moghaddam, Y.~Weiss, and S.~Avidan, ``Spectral bounds for sparse pca: Exact
  and greedy algorithms,'' in \emph{Advances in neural information processing
  systems}, 2005, pp. 915--922.

\bibitem{johnstone2009consistency}
I.~M. Johnstone and A.~Y. Lu, ``On consistency and sparsity for principal
  components analysis in high dimensions,'' \emph{Journal of the American
  Statistical Association}, vol. 104, no. 486, 2009.

\bibitem{d2008optimal}
A.~d'Aspremont, F.~Bach, and L.~E. Ghaoui, ``Optimal solutions for sparse
  principal component analysis,'' \emph{The Journal of Machine Learning
  Research}, vol.~9, pp. 1269--1294, 2008.

\bibitem{krauthgamer2013semidefinite}
R.~Krauthgamer, B.~Nadler, and D.~Vilenchik, ``Do semidefinite relaxations
  really solve sparse pca?'' \emph{arXiv preprint arXiv:1306.3690}, 2013.

\bibitem{deshpande2013sparse}
Y.~Deshpande and A.~Montanari, ``Sparse pca via covariance thresholding,''
  \emph{arXiv preprint arXiv:1311.5179}, 2013.

\bibitem{amini2009high}
A.~A. Amini and M.~J. Wainwright, ``High-dimensional analysis of semidefinite
  relaxations for sparse principal components,'' \emph{Annals of statistics},
  vol.~37, no.~5, pp. 2877--2921, 2009.

\bibitem{rangan2012iterative}
S.~Rangan and A.~K. Fletcher, ``Iterative estimation of constrained rank-one
  matrices in noise,'' in \emph{Information Theory Proceedings (ISIT), 2012
  IEEE International Symposium on}.\hskip 1em plus 0.5em minus 0.4em\relax
  IEEE, 2012, pp. 1246--1250.

\bibitem{matsushita2013low}
R.~Matsushita and T.~Tanaka, ``Low-rank matrix reconstruction and clustering
  via approximate message passing,'' in \emph{Advances in Neural Information
  Processing Systems}, 2013, pp. 917--925.

\bibitem{berthet2013computational}
Q.~Berthet and P.~Rigollet, ``Computational lower bounds for sparse pca,''
  \emph{arXiv preprint arXiv:1304.0828}, 2013.

\bibitem{kabashima2014phase}
Y.~Kabashima, F.~Krzakala, M.~M{\'e}zard, A.~Sakata, and L.~Zdeborov{\'a},
  ``Phase transitions and sample complexity in bayes-optimal matrix
  factorization,'' \emph{arXiv preprint arXiv:1402.1298}, 2014.

\bibitem{YedidiaFreeman03}
J.~Yedidia, W.~Freeman, and Y.~Weiss, ``Understanding belief propagation and
  its generalizations,'' in \emph{Exploring Artificial Intelligence in the New
  Millennium}.\hskip 1em plus 0.5em minus 0.4em\relax San Francisco, USA:
  Morgan Kaufmann, 2003, pp. 239--236.

\bibitem{donoho2012information}
D.~L. Donoho, A.~Javanmard, and A.~Montanari, ``Information-theoretically
  optimal compressed sensing via spatial coupling and approximate message
  passing,'' \emph{IEEE transactions on information theory}, vol.~59, no.~11,
  pp. 7434--7464, 2013.

\bibitem{hoyle2004principal}
D.~C. Hoyle and M.~Rattray, ``Principal-component-analysis eigenvalue spectra
  from data with symmetry-breaking structure,'' \emph{Physical Review E},
  vol.~69, no.~2, p. 026124, 2004.

\bibitem{baik2005phase}
J.~Baik, G.~Ben~Arous, and S.~P{\'e}ch{\'e}, ``Phase transition of the largest
  eigenvalue for nonnull complex sample covariance matrices,'' \emph{Annals of
  Probability}, pp. 1643--1697, 2005.

\end{thebibliography}
\end{document}